\documentclass[unnumsec,webpdf,contemporary,large]{oup-authoring-template}%

\graphicspath{{Fig/}}
\usepackage{svg}

\theoremstyle{thmstyleone}%
\theoremstyle{thmstyletwo}%
\theoremstyle{thmstylethree}%

\begin{document}

\journaltitle{Journal Title Here}
\DOI{DOI HERE}
\copyrightyear{2022}
\pubyear{2019}
\access{Advance Access Publication Date: Day Month Year}
\appnotes{Paper}

\firstpage{1}


\title[PyamilySeq]{PyamilySeq: A Python Tool for Interpretable Gene (Re)Clustering and Pangenomic Inference Across Species and Genera}

\author[1,2,3,4,5$\ast$]{Nicholas J Dimonaco}

\authormark{Nicholas J Dimonaco}

\address{
$^{\text{\sf 1}}$Department of Medicine, McMaster University, Hamilton, ON, Canada\\ 
$^{\text{\sf 2}}$Farncombe Family Digestive Health Research Institute, McMaster University, Hamilton, ON, Canada\\
$^{\text{\sf 3}}$School of Biological Sciences, Queen’s University Belfast, Belfast, BT9 5DL, Northern Ireland, UK\\
$^{\text{\sf 4}}$Institute for Global Food Security, Queen's University Belfast, Belfast, BT9 5DL, UK\\
$^{\text{\sf 5}}$Department of Computer Science, Aberystwyth University, Aberystwyth, SY23 3DB, Wales, UK\\ 
}

\corresp[$\ast$]{Corresponding author. \href{email:nicholas@dimonaco.co.uk}{nicholas@dimonaco.co.uk}}

\received{Date}{0}{Year}
\revised{Date}{0}{Year}
\accepted{Date}{0}{Year}



\abstract{
PyamilySeq is a Python-based tool designed for interpretable gene clustering and pangenomic inference, supporting analyses at both species and genus levels.
It facilitates the clustering of gene sequences into families based on sequence similarity using CD-HIT, and can take the output of tried-and-tested sequence clustering tools such as CD-HIT, BLAST, DIAMOND, and MMseqs2. 
PyamilySeq is distinctive in its ability to integrate new sequences into existing clusters, providing a robust framework for iterative analysis while preserving the original clusters, useful when reannotating genomes.
In addition to the standard `Species' mode which as with other tools performs core-gene analysis across a species range, PyamilySeq can be run in `Genus' mode where it detects the presence of gene `families' shared across multiple genera.
These features enhance the tool's applicability for ongoing and past genomic studies and comparative analyses. 
PyamilySeq generates comprehensive outputs, including gene presence-absence matrices and aligned sequence data, enabling downstream analysis and interpretation of the identified gene groups and pangenomic data.
}


\maketitle

\section{Introduction}

Gene clustering, the \textit{in silico} process of grouping genes by a common characteristic such as function or sequence similarity, is essential for understanding the genetic diversity and functional capabilities of groups of microorganisms. 
Existing tools often pose challenges in terms of usability, applicability and interoperability. 
PyamilySeq addresses these challenges by providing a user-friendly platform for clustering gene sequences into families based on sequence similarity, supporting both species-level and genus-level analyses. 
This toolkit extends the functionalities of the gene family/pangenome tool used in the StORF-Reporter genome annotation publication \cite{dimonaco2023storf}, offering improved performance, features and ease of use.

One of PyamilySeqs advantages over other tools is its simplicity and level of user control. 
For example, not only can users perform the clustering of gene sequences externally to PyamilySeq if they wish, they can practically use any method, and thus clustering parameters, as PyamilySeq can read in any edge\-list file (\textit{Node1 Node2}).
PyamilySeq can also take DNA or Amino Acid sequences/clusters as input.
Additionally, PyamilySeq provides the option to specify the `core/soft-core/accessory' definitions so that they are more aligned to the input data distribution.

PyamilySeq is available at \url{https://pypi.org/project/PyamilySeq/} and can be installed with \textit{pip install PyamilySeq} on most systems running Python3.6 or later.
The full user-menu and test data can be found at GitHub (\url{https://github.com/NickJD/PyamilySeq}).

\section{Methods}

\subsection{Seq\_Combiner}
A major hurdle in bioinformatic analysis is file format compatibility. 
Tools such as Panaroo \cite{tonkin2020producing} are picky and require the GFF files that the user wants processed in a particular format which can often cause run-time errors or crashes.
As part of the PyamilySeq toolkit, Seq\_Combiner is provided to allow for the generation of compatibility-assured input files for PyamilySeq.
Seq\_Combiner takes as input a directory of either matching FASTA and GFF files or GFF files with `\#\#FASTA - Seq' appended to the bottom and appends the name of the file to the start of each gene ID.
\newpage
An example of this process is shown below:

\begin{verbatim}
GFF Filename: Escherichia_coli_110957.ASM48561v1.gff
GeneID: E1CxwDWD5uDlWpi

>Escherichia_coli_110957.ASM48561v1|E1CxwDWD5uDlWpi
>Escherichia - [GenusID]
>Escherichia_coli_110957.ASM48561v1 - [SpeciesID]
>Escherichia_coli_110957.ASM48561v1|E1CxwDWD5uDlWpi - [GeneID]
\end{verbatim}

\subsection{Overview}
\begin{figure}[t]
\includegraphics[width=0.45\textwidth,height=0.6\textheight,keepaspectratio]{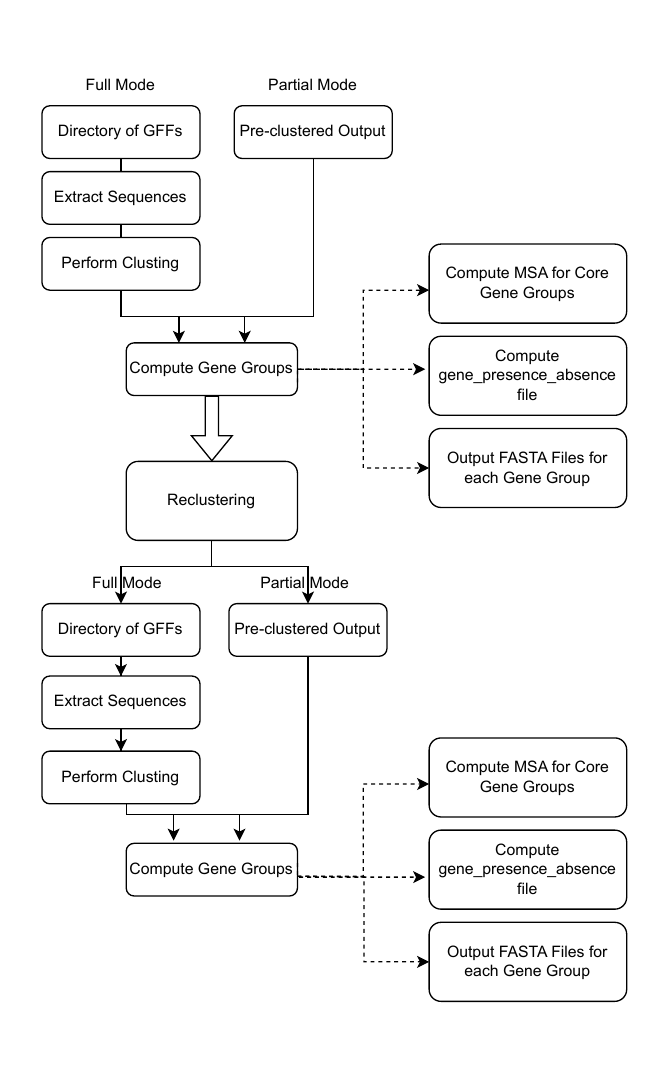}
  \caption{Overview of Full and Partial run modes}
  \label{fig:Overview}
\end{figure}

PyamilySeq is implemented in Python and requires Python 3.6 or higher. 
It supports multiple sequence clustering tools, including CD-HIT \cite{fu2012cd}, BLAST \cite{altschul1990basic}, DIAMOND \cite{buchfink2021sensitive}, and MMseqs2 \cite{steinegger2017mmseqs2}, and it can process input files in combined or separate GFF+FASTA formats. 
PyamilySeq operates in two primary modes: Full Mode and Partial Mode - see Figure \ref{fig:Overview}.

\subsection{Full Mode}
In Full Mode, PyamilySeq processes a directory of GFF+FASTA files, performs sequence clustering using the specified tool, and generates a gene presence-absence matrix. The workflow involves:
\begin{enumerate}
    \item Reading and combining input sequences if necessary.
    \item Clustering sequences using the chosen algorithm.
    \item Identify and reports gene groups based on user-parameters.
    \item Generating a gene presence-absence matrix formatted for downstream analysis.
    \item Optionally aligning representative `core' sequences using MAFFT \cite{katoh2002mafft} and outputting concatenated aligned sequences.
\end{enumerate}


\subsection{Partial Mode}

Partial Mode allows the user to provide PyamilySeq with a pre-clustered set of genes in either CD-HIT or edge-list formats and then PyamilySeq only performs steps 3-5 of Full Mode.

\subsubsection{Recluster}

\begin{figure}[h]
\includegraphics[width=0.5\textwidth,height=0.6\textheight,keepaspectratio]{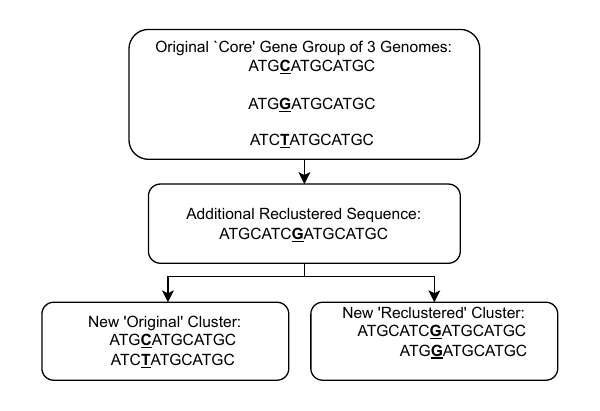}
  \caption{This figure shows how adding additional sequences can break original clusters and thus decrease the overall number of core/soft-core gene groups.}
  \label{fig:Reclustered}
\end{figure}

\begin{figure}[h]
\includegraphics[width=0.5\textwidth,height=0.6\textheight,keepaspectratio]{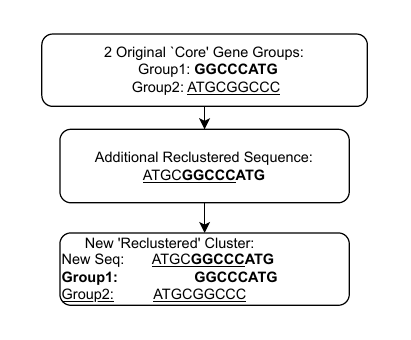}
  \caption{This figure shows how adding additional sequences can combine two or more original clusters and thus change the overall number of core/soft-core gene groups.}
  \label{fig:Reclustered_Combined}
\end{figure}

As part of the Partial Mode operation, PyamilySeq can also recluster additional sequences with a pre-clustered output.
This was performed in the StORF-Reporter publication \cite{dimonaco2023storf} as StORFs which are often longer than their 'pre-found' counterparts can actually break up or combine gene groups as can be seen in Figures \ref{fig:Reclustered} and \ref{fig:Reclustered_Combined}.
Reclustering can be a very powerful tool to see where novel gene predictions lay in relation to a canonical annotation and clustering analysis.

\subsection{Group Mode}

\begin{figure}[h]
\includegraphics[width=0.5\textwidth,height=0.6\textheight,keepaspectratio]{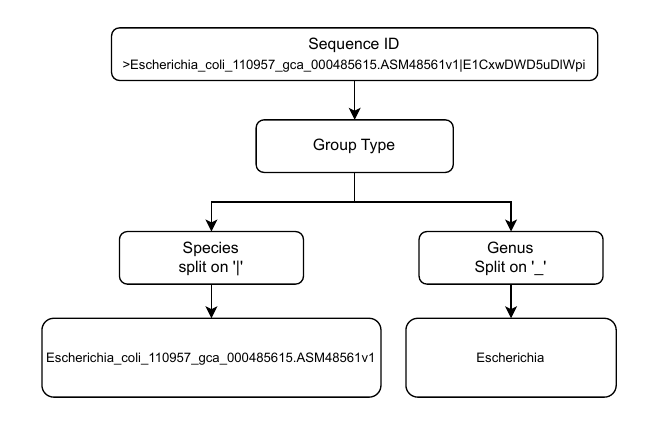}
  \caption{Overview of Group mode}
  \label{fig:Group_Mode}
\end{figure}

Group mode allows PyamilySeq to either perform gene group identification in `Species' or `Genus' mode. 
As can be seen in Figure \ref{fig:Group_Mode}, the two modes harness the sequence IDs to identify the species or genus from which the sequence is from. 
Using this information, in Species mode, PyamilySeq will perform as other contemporary pangenome tools such as Roary \cite{page2015roary} and Panaroo \cite{tonkin2020producing} do and identifies gene groups such as 'core', 'soft-core' and 'accessory'.
However, in Genus mode, PyamilySeq instead identifies gene groups that are shared across different genera and thus allows for a much broader overview of a genes presence across disparate taxa.

\subsection{Output}

PyamilySeq can provide different outputs depending on user requirements.
Firstly, the \textit{summary\_statistics.txt} file is always produced and broadly follows the format as reported by Roary and Panaroo albeit with some minor changes (\# notates the percentage used to separate each group - defaults being 99, 95, 15, 0).
Here `First' and `Second' is used to notate the results of either the initial or reclustered clustering, respectively.

The below definitions broadly follow those used in the StORF-Reporter paper.
\begin{itemize}
\item \textbf{First\_core\_\#}: This group broadly follows the definition setout by Roary and Panaroo and reports gene groups from the First round of clustering.
\item \textbf{extended\_core\_\#}: This group reports the number of gene groups that were extended into the notated groups via the addition of reclustered sequences added during the second round of clustering. 
\item \textbf{combined\_core\_\#}: This group reports the number of clusters where reclustered sequences combined at least 2 or more original cluster representatives together as seen in Figure \ref{fig:Reclustered_Combined}.
\item \textbf{Second\_core\_\#}: This group refers to the clusters grouped with only the reclustered sequences being counted.
\item \textbf{only\_Second\_core\_\#} This group refers to the clusters which solely contain reclustered sequences. 
\end{itemize}

\section{Results}
\subsection{Case Study: Pangenomic Analysis of \textit{Escherichia coli}}
PyamilySeq was applied to a dataset consisting of 10 \textit{Escherichia coli (E. coli)} genomes to evaluate its performance and utility. 
For sequence clustering, CD-HIT was used with a sequence identity threshold of 0.90 and a length difference threshold of 0.60.
Pangenome analysis of the same set of 10 genomes was also conducted using both Panaroo and Roary - The results are below.

\paragraph{PyamilySeq:}
\begin{verbatim}
    Gene Groups:
    First_core_99: 3099
    First_core_95: 0
    First_core_15: 3200
    First_core_0: 4996
    Total Number of Gene Groups (Including Singletons): 11295
\end{verbatim}

\paragraph{Panaroo:}
\begin{verbatim}
    Core genes	(99% <= strains <= 100%)	3309
    Soft core genes	(95% <= strains < 99%)	0
    Shell genes	(15% <= strains < 95%)	3027
    Cloud genes	(0% <= strains < 15%)	2607
    Total genes	(0% <= strains <= 100%)	8943
\end{verbatim}

\paragraph{Roary:}
\begin{verbatim}
    Core genes	(99% <= strains <= 100%)	3076
    Soft core genes	(95% <= strains < 99%)	0
    Shell genes	(15% <= strains < 95%)	3458
    Cloud genes	(0% <= strains < 15%)	6188
    Total genes	(0% <= strains <= 100%)	12722
\end{verbatim}

\begin{figure}[h]
\includegraphics[width=0.5\textwidth,height=0.5\textheight,keepaspectratio]{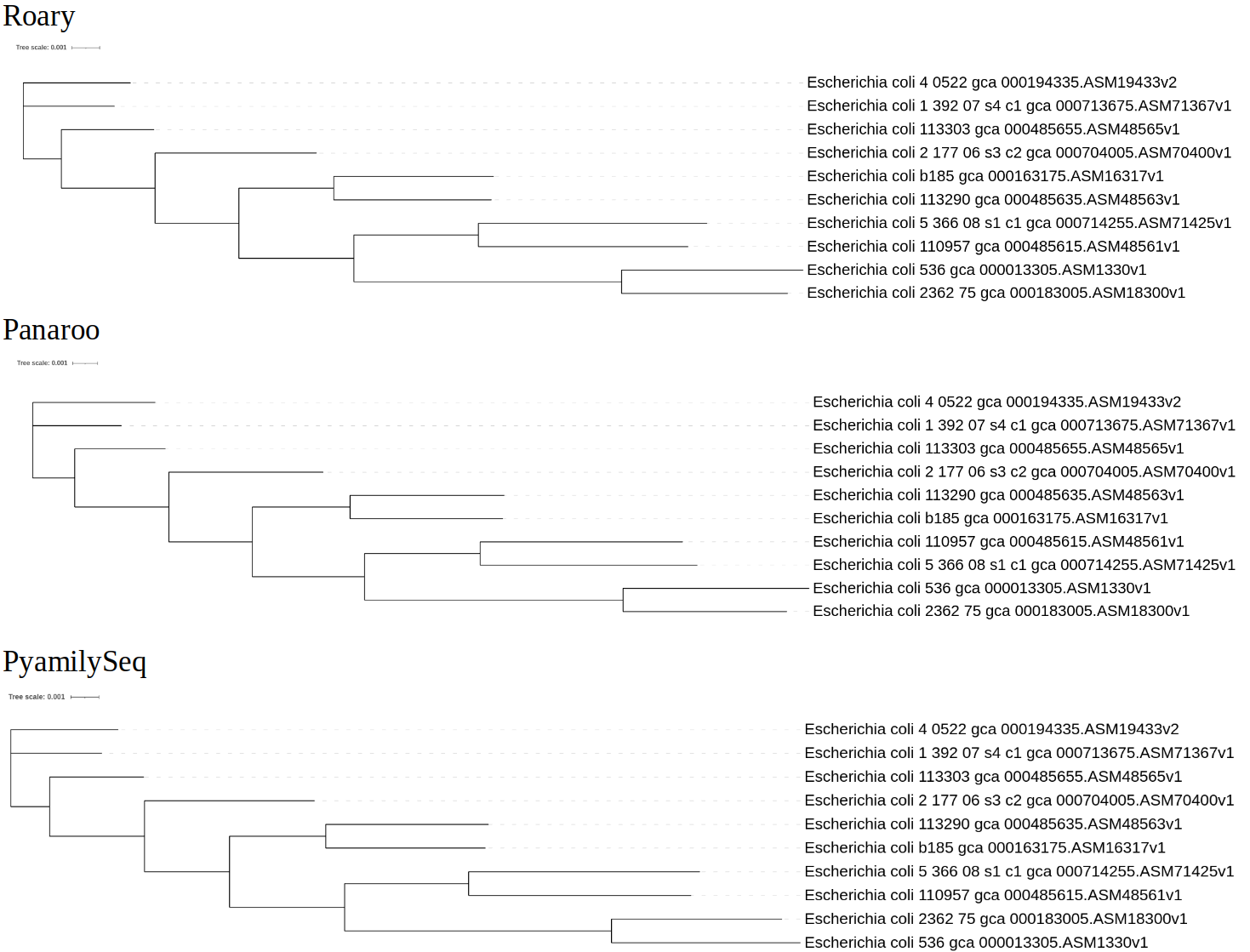}
  \caption{Three phylogenetic trees built using the `core-gene' families identified by Roary, Panaroo and PyamilySeq.}
  \label{fig:PhyloTree}
\end{figure}

All three tools can produce a `core-gene' multiple sequence alignment (MSA) that can be used for phylogenetic tree construction.
PyamilySeq uses MAFFT \cite{katoh2002mafft} to align each gene group independently and then concatenate the resulting alignments together, mostly following Roary and Panaroo.
Below are three phylogenetic trees built using FastTree2 \cite{price2010fasttree} from the MSAs from Roary, Panaroo and PyamilySeq, respectively.
As can be seen in Figure \ref{fig:PhyloTree}, there are no structural differences between the trees.

\subsubsection{Reclustering Capabilities}
PyamilySeq's ability to recluster new sequences into existing clusters was demonstrated by adding sequences from a new gene annotation tool, namely StORF-Reporter. 
The updated clusters maintained the integrity of the original clusters while integrating the new sequences, showcasing the tool's iterative analysis capability.
\paragraph{Reclustered StORF-Reporter Predictions:}

\begin{verbatim}
    Gene Groups:
    First_core_99: 2914
    First_core_95: 0
    First_core_15: 3352
    First_core_0: 5647
    extended_core_99: 0
    extended_core_95: 0
    extended_core_15: 0
    extended_core_0: 0
    combined_core_99: 0
    combined_core_95: 0
    combined_core_15: 0
    combined_core_0: 0
    Second_core_99: 0
    Second_core_95: 0
    Second_core_15: 5
    Second_core_0: 18
    only_Second_core_99: 0
    only_Second_core_95: 0
    only_Second_core_15: 1905
    only_Second_core_0: 4140
    Total Number of Gene Groups (Including Singletons): 11913
\end{verbatim}

\subsection{Case Study: Cross-Genus Analysis of \textit{Escherichia coli}, \textit{Bacillus subtilis}, \textit{Caulobacter vibrioides}, \textit{Mycoplasma genitalium}, \textit{Pseudomonas aeruginosa} and \textit{Staphylococcus aureus}}

A unique aspect of PyamilySeq is its ability to identify gene groups shared across different genera.
Below is the result of running PyamilySeq in Partial and Genus mode on two of each aforementioned genomes. 
While contemporary pangenomic tools would report no gene groups/families as being core, with Genus mode, we are able to see that there are a small number gene groups present across more than one genus. 
\begin{verbatim}
    Genus Groups:
    First_genera_1:	28549
    Second_genera_1:	249
    only_Second_genera_1:	17551
    First_genera_2:	12
    Second_genera_2:	0
    only_Second_genera_2:	0
    First_genera_>:	0
    Second_genera_>:	0
    only_Second_genera_>:	0
    Total Number of Gene Groups (Including Singletons): 28561
\end{verbatim}

\section{Discussion and Conclusion}
PyamilySeq addresses several limitations of existing post- gene clustering and pangenomic analysis tools. 
Its user-friendly interface and flexible input options make it accessible to researchers with varying levels of computational expertise. 
The ability to integrate new sequences post-clustering is a significant advantage, allowing for iterative analysis as new data and evidence becomes available.
The additional capability of clustering at the genus-level provides a new approach to study genes shared across disparate taxa.

Although PyamilySeq is a powerful and user-friendly tool for gene clustering and pangenomic analysis, it is not designed to overtake contemporary techniques but instead provides avenue for research.

\section{Software versions used}
\begin{verbatim}
    PyamilySeq: v0.6.0
    CD-HIT: v4.8.1
    Panaroo: v1.5.0
    Roary: v3.13.0
    MAFFT: v7.490 
    Fasttree: v2.1.11
\end{verbatim}

\section{Competing interests}
No competing interest is declared.



\bibliographystyle{plain}
\bibliography{main}

\end{document}